\documentclass[useAMS,usenatbib,letters]{mn2e}

\usepackage{psfig}
\usepackage{amssymb}    

\newcommand\name{LEDA 135736}   
\newcommand\oxy{[OIII] }           

\title[Mapping Seyfert galaxy \name] {Highly ionized gas on galaxy
scales: mapping the interacting Seyfert galaxy \name
\thanks{Based on observations made with ESO Telescopes at the Paranal
  Observatory under programme 078.B-0194(A).}}

\author[J.Gerssen]{J.~Gerssen$^{1}$\thanks{E-mail: jgerssen@aip.de},
D.~J.~Wilman$^{2}$, L.~Christensen$^{3}$, R.~G.~Bower$^{4}$ and
V. Wild$^{5}$\\
$^{1}$Astrophysikalisches Institut Potsdam, An der Sternwarte 16,
      14482 Potsdam, Germany.\\
$^{2}$Max-Planck-Institut f\"ur Extraterrestrische Physik,
      Giessenbachstrasse, 85748 Garching, Germany.\\
$^{3}$European Southern Observatory, Casilla 19001, Santiago 19, Chile.\\
$^{4}$Durham University, Department of Physics, Durham, DH1 3LE,, UK.\\
$^{5}$Max-Planck-Institut f\"ur Astrophysik,
Karl-Schwarzschild-Str. 185741 Garching, Germany}

\begin{document}

\date{Accepted 2008 November 6}


\maketitle

\label{firstpage}

\begin{abstract}
We have used the VIMOS IFU to map the properties of the Seyfert 1.9
galaxy \name.  These maps reveal a number of interesting features
including: an Extended Narrow Line Region detectable out to
$\sim9$~kpc, an area of intense star formation located at a projected
distance of 12 kpc from the centre, an elliptical companion galaxy,
and kinematic features, aligned along the long-axis of the ENLR, that
are consistent with radio jet-driven mass outflow.  We propose that
the ENLR results from extra-planar gas ionized by the AGN, and that
the AGN in turn might be triggered by interaction with the companion
galaxy, which can also explain the burst of star formation and
morphological features.  Only about two percent of the ENLR's kinetic
energy is in the mass outflow.  We infer from this that the bulk of
mechanical energy imparted by the jet is used to heat this gas.
\end{abstract}



\section{Introduction}

In the hierarchical picture of galaxy evolution, the formation and
growth is driven by mergers. Such events affect galaxies on all scales
down to their nuclei where they can trigger AGN activity
(e.g. Springel et al., 2005; di~Matteo et al., 2005; Cattaneo et
al. 2005).  Observationally, there is some evidence for a link between
AGN and mergers (e.g. Sanchez et al., 2005; Kuo et al., 2008). But
this connection is difficult to establish in general as the brightest
AGN typically reside at redshifts too large to determine the properties
of the host galaxy and its neighbours.

It is increasingly recognized, however, that feedback generated by the
AGN itself plays a crucial role in galaxy evolution.  A jet powered by
an AGN can drive large amounts of material out of the host system and
thus significantly affect its subsequent evolution. Alternatively the
AGN might primarily act on the galaxy halo, heating the halo, reducing
the net cooling rate and possibly expelling some material.  These two
modes of feedback are often referred to as the ``quasar-mode'' and
``radio mode'' respectively.  Models that incorporate AGN feedback
can, for instance, account for the observed galaxy luminosity
functions (Croton et al.\ 2006, Bower et al. 2006) using the ``radio
mode'' and can reproduce the observed $M_{\rm BH}$-$\sigma$
correlation (e.g. di~Matteo et al. 2005) primarily using the ``quasar
mode''.  Observational evidence of quasar-mode feedback is emerging
for in intrinsically bright AGN such as QSOs (Letawe et al.,2008) and
compact radio sources (Nesvadba et al., 2007; Holt et al., 2008).
Meanwhile observations of X-ray cavities associated with radio sources
in galaxy clusters provide clear evidence of the effectiveness of the
``radio mode'' (Birzan et al.\ 2004, Allen et al.\ 2006). A key issue
is to observationally determine the relative importance of these two
modes in lower mass systems.
In this letter we present VIMOS IFU observations of a nearby, low
luminosity AGN, the Seyfert 1.9 galaxy \name\ at a redshift of
$z=0.066$.  The AGN in this system ionizes an Extended Narrow Line
Region (ENLR) up to at least 9 kpc.  Interaction with an elliptical
companion, at a projected distance of 11.6 kpc, is probably triggering
jet activity (as well as off-centre star formation) that we observe
indirectly in the H$\alpha$ and [OIII] kinematic maps.

\section{Data Reduction and Analysis}

\name\ was observed with the VIMOS Integral Field Unit (IFU) as part
of a project to map the properties of a sample of 24 galaxies selected
randomly from the Sloan Digital Sky Survey (SDSS; York et al. 2000).
All data were obtained using the medium resolution setup (wavelength
range: 5000 - 9000 \AA, dispersion, 2.5 \AA/pix) covering a
field-of-view of 27x27 arcsec (0.67 arcsec/spaxel).  We obtained two
30 minute exposures on this galaxy during service mode observations in
January 2007 (in seeing conditions of about 1.5 arcsec).  A detailed
description of the data reduction will be given in a forthcoming paper
(Gerssen et al, 2008).  Briefly, we used the ESO VIMOS pipeline to
perform the basic reduction steps up to spectrum extraction and
wavelength calibration.  The post-processing steps (e.g. throughput
correction, flux calibration, and exposure combination) to create the
final data cube ($x$,$y$,$\lambda$) were performed using custom
written IDL scripts.

To analyse the emission line data, we independently fit the
H$\alpha$+[NII] group, the [OIII] doublet, the [SII] doublet and the
H$\beta$ emission line.  Each line is fit with a single Gaussian, and
for each set of lines the relative position and widths of each line
are fixed to each other as they trace the same kinematics.  For
example, in a three component fit to the H$\alpha$+[NII] emission
lines we tie the centroids and line widths to the H$\alpha$ line.  In
this case there are six free parameters: the amplitudes of the three
emission lines, the line centroid and line width, and a constant
continuum level.  We do not include an additional broad H$\alpha$
component in the emission line analysis. The Broad Line Region (BLR)
in \name\ is only detectable in the H$\alpha$ line and then only in
spectra close to the nucleus, where it is so broad as to have no
influence on the fit.

\section{Results}

The galaxy \name\ is a Seyfert 1.9 at a distance\footnote{The
luminosity distance is calculated assuming: $\Omega_\Lambda$=$0.73$,
$\Omega_M$=$0.27$, $H_0$=$71$ km s$^{-1}$ Mpc$^{-1}$. The scale is
$1.25$ kpc/$\prime\prime$.} of 293 Mpc.  Its basic properties are
listed in Table~1.  It stands out in our sample because the radial
dependence of its strongest emission lines (H$\alpha$, H$\beta$,
[NII]6584 [OIII]5007) indicates a high ionization state out to large
radii, see Figure~\ref{f:bpt}.  In this so-called BPT diagram (Baldwin
et al. 1981) we plot results derived using a synthetic
annulus-aperture (2 arcsec width) of increasing radius.  Remarkably,
\name\ is located on the AGN `wing' of the BPT diagram out to a radius
of at least 9 kpc. This implies a role for strong ionizing radiation,
probably associated with the AGN, on galaxy-wide scales.


\begin{table}
  \centering
    \caption{Properties of \name}
    \begin{tabular}{@{}lc@{}}
    \hline

    \multicolumn{2}{c}{\it Basic} \\
    RA (J2000)  & 09h59m39.8s  \\
    Dec (J2000) & +00d35m14s   \\
    Type        & Sy 1.9       \\
    $z$         & $0.066$      \\

    \multicolumn{2}{c}{\it Luminosity} \\
    $L_{\rm X}$ (erg s$^{-1}$)         & $7.2 \times 10^{42}$   \\
    $L_{\rm [OIII]}$ (erg s$^{-1}$) & $4.2 \times 10^{40}$     \\ 
    $L_{{\rm H}\alpha}$ (erg s$^{-1}$) & $4 \times 10^{40}  $   \\ 
    $L_{\rm 1.4GHz}$ (erg s$^{-1}$ Hz$^{-1}$)      & $7.7 \times 10^{29}$   \\

    \hline
    \end{tabular}
    \label{t:properties}
\end{table}


\subsection{Morphology}

\name\ displays a complex morphological structure that is strongly
wavelength dependent.  A composite colour image derived from our data
cube is shown in panel (a) of Figure~\ref{f:mosaic}.  The two
brightest knots, labelled A and B, coincide respectively with the
nucleus of the host and that of a nearby galaxy.  Their projected
separation is 11.6 kpc.  Hence, it is likely that the two systems are
interacting.  A clear manifestation of interaction are the faint knots
visible NE of the host nucleus. The knot labelled C coincides with the
peak of a resolved ultraviolet source (GALEX database).  The H$\alpha$
and [OIII] line flux maps (panels d and g) respectively) do show
prominent features in region C as well.  Interestingly, the H$\alpha$
peak intensity is somewhat stronger for the off-centre peak than on
the nucleus itself.  The average line ratios over region C (shown in
Figure~1) are consistent with ionization by young stars.  This area is
likely associated with off-centre star formation (at $\sim 10.8$
projected kpc from the nucleus) triggered by interaction with the
companion galaxy.

As the companion galaxy shows no emission lines we establish its
nature using a near infrared H-band image obtained from the UKIRT
Infrared Deep Sky Survey (UKIDSS; Lawrence et al. 2007).  The surface
brightness profile (derived with ELLIPSE in IRAF) is consistent with
the light profile of an elliptical galaxy with an effective radius of
$R_{\rm eff} \sim 2.5$ kpc.

\subsection{Stellar kinematics}

To constrain the systemic velocities of the host and the companion we
sum the spectra in our data cube over regions A and B.  We use the
pixel-fitting method of (Cappellari \& Emsellem 2004) to fit the
summed specta with a set of stellar template spectra observed with
EMMI on the NTT (convolved to the VIMOS instrumental resolution of 7.3
\AA). The comparison, shown in panel (b), between the stellar
absorption line spectra extracted at locations A and B demonstrates
that the companion galaxy is also close in velocity space ($\Delta
v_{\rm los} \sim 600$ km~s$^{-1}$).

\subsection{Extended Narrow Line Region}

\begin{figure}
   \centerline{\psfig{figure=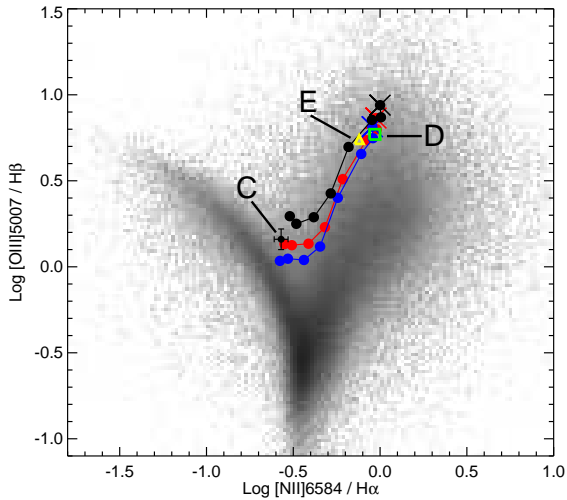,width=0.45\textwidth}}
   \caption{ To broadly quantify the radial dependence of the
    properties of \name\ we measure emission line fluxes in annuli of
    increasing radius.  The result is shown as a `track' (black line)
    on top of a BPT diagram derived from single-aperture measurements
    of some 500,000 galaxies in the SDSS database (Brichmann et
    al. 2004).  Our central aperture (1$^{\prime\prime}$ radius) is
    indicated by the cross.  The radius increases by
    2$^{\prime\prime}$ for successive points along the track.
    Standard corrections for Balmer absorption of EW=2 \AA \ and EW=4
    \AA \ (red and blue coloured tracks respectively) do not
    significantly change the results.  The location of the off-centre
    star forming region (region C, see section 3.1) in this diagram is
    marked by the black square with error bars and is corrected for
    stellar absorption (EW=2 \AA).  The $1-\sigma$ errors on the
    azimuthally-averaged measurements are smaller than the symbol
    sizes.  The open square and triangle show the line ratios in box D
    \& E respectively (section~4).  }
   \label{f:bpt}
\end{figure}

\begin{figure*}
   \centerline{\psfig{figure=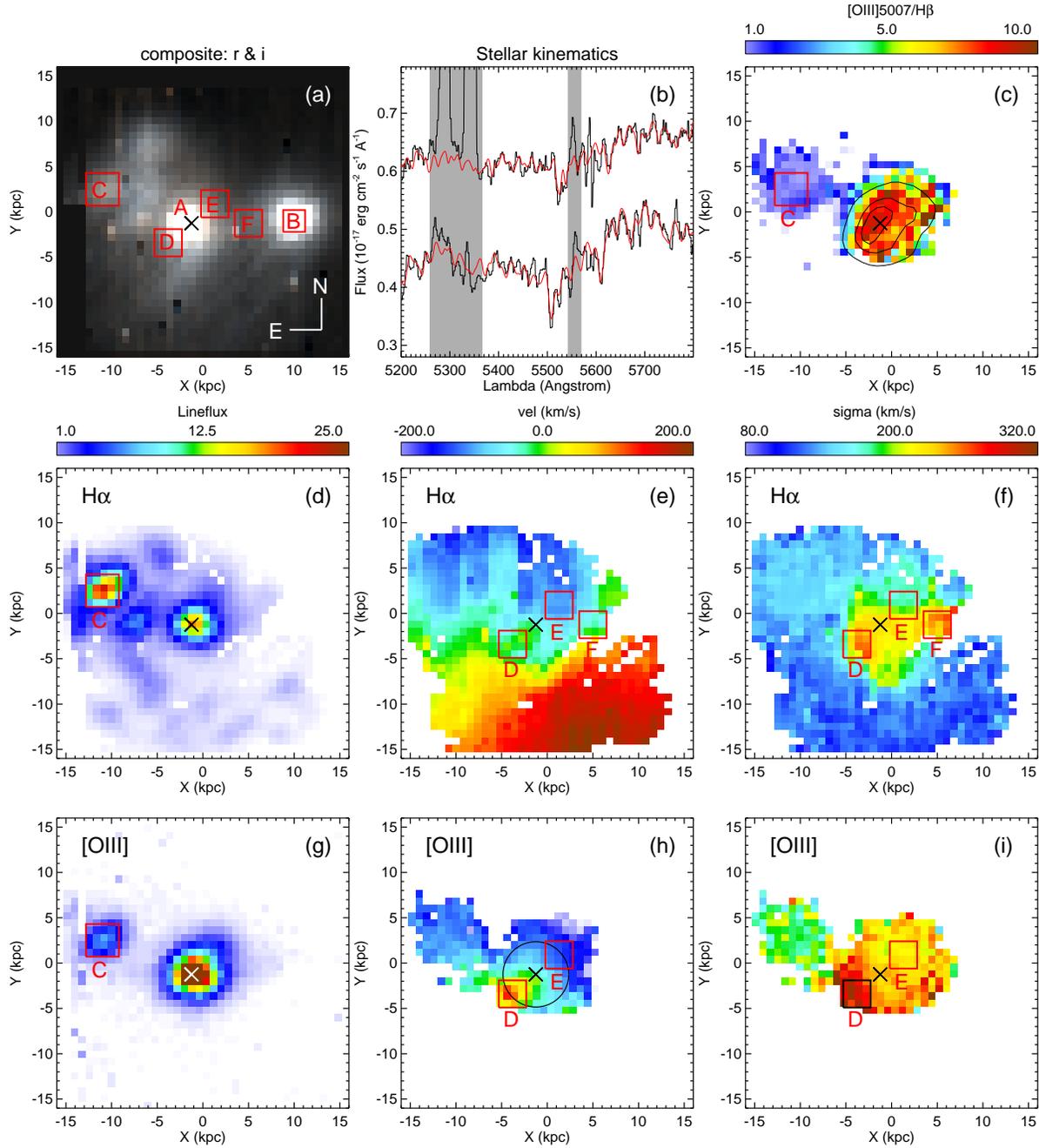,width=0.9\textwidth}}
   \caption{Properties of \name\ derived from the VIMOS IFU data.  In
     all panels the orientation is such that North is up and East is
     the left. The position of the Broad Line Region, and hence the
     nucleus itself, is marked with the cross in each panel.  {\bf Top
     row:} {\bf a:} A composite image created by emulating the SDSS
     r-and i-band filter profiles. The labelled regions (A to F) are
     discussed in detail in the text.  {\bf b:} Absorption line
     spectra of the companion galaxy (bottom curve) summed over the
     box labelled B in panel (a). The top curve shows the spectrum of
     the host galaxy summed over a similar sized area.  The spectra
     are fit over a wavelength range around the Mgb stellar absorption
     line feature ($\lambda\lambda \sim 5170$ \AA) with single stellar
     template spectra. The best fits are shown in red. The wavelength
     contaminated by emission lines are excluded in this fit (dashed
     regions).  {\bf c:} Ionization map.  The high values $f_{{\rm
     [OIII]}\lambda5007}/f_{{\rm H}\beta}>>1$ suggest an AGN as the
     source of ionization. The orientation of this Extended Narrow
     Line Region is highlighted by the overplotted contours.  Its long
     axis lies along a position angle of $-45^\circ$.  In this and
     subsequent panels the colour map scales are linear. Results are
     shown only for spaxels with lineflux S/N ratios $\ge 3$.  {\bf
     Middle row:} Maps of the H$\alpha$ line flux, the H$\alpha$
     velocity field, and the velocity dispersions. The labelled boxes
     (C: off-centre SF region, D: \& E: off-centre velocity dispersion
     peaks) are discussed in detail in the text. {\bf Bottom row:}
     Similar to the middle row but showing the [OIII]5007 properties
     on the same scales as the H$\alpha$ properties. The black circle
     in panel(h) delineates the upper limit on the size of the
     unresolved radio source.
   \label{f:mosaic}}
 \end{figure*}

To examine the result shown in Figure~\ref{f:bpt} in more detail we
can use our data cube to derive the line ratios in each spatial
element individually.  The full 2-D line ratio map of
[OIII]$\lambda5007$/H$\beta$ is shown in panel (c) of
Figure~\ref{f:mosaic}.  Consistent with our azimuthally averaged
result (Fig.~1), the map shows an extended region of highly ionized
gas.  The observed line ratios require a very strong ionization field
(hard UV spectrum) and are typical for the Narrow Line Region (NLR) of
a Seyfert galaxy. Such highly ionized gas, not confined to the nuclear
regions of a galaxy, is known as an Extended Narrow Line Region (ENLR)
(e.g. Bennert et al., 2006).  As the overplotted contours demonstrate,
the spatially resolved part ($\lesssim$5~kpc) of the ENLR in \name\ is
somewhat elongated ($a/b>2$) with the long axis at a PA $\approx
-45^\circ$.

Panels (d) and (e) of Figure~\ref{f:mosaic} respectively show the
H$\alpha$ flux and velocity field. The Balmer lines are relatively
more sensitive to less strongly ionized radiation fields than the
[OIII] line (as utilised by the BPT diagram) and so more closely
traces gas ionized by young stars.  The rotating disk traced by the
H$\alpha$ line is therefore probably unrelated to the ENLR, with gas
ionized by normal star formation in the disk.  Our radially-averaged
results (Fig.~1) illustrate that out to $\sim 9$ kpc the [OIII] line
includes contributions from both disk gas ionized by young stars, and
the more highly ionoized gas from the AGN (the ENLR).  Beyond $\sim 9$
kpc (including region~C) star formation dominates, and the line ratios
are consistent with normally star-forming galaxies.

\subsection{Gas kinematics}

The gas kinematics derived from the H$\alpha$ and [OIII]5007 emission
lines are shown in Figure~\ref{f:mosaic} along the middle row and
bottom row respectively.  The average error on the best-fit (as
described in section~2) line centroids and line widths is $\sim 10$ km
s$^{-1}$.

On a global scale the H$\alpha$ velocity field (panel e) is consistent
with a simple rotating disk model. Near the systemic velocity,
however, the data are significantly distorted from the straight line
zero velocity contour (green in our map) predicted by this model.  The
peak in the H$\alpha$ velocity dispersion map (panel f) does not
coincide with the nucleus itself but is located SE of it along the
long-axis of the ENLR in a region that we label D.  Interestingly, a
second dispersion peak, labelled F, is visible in the direction of the
companion galaxy.  The off-centre dispersion peaks cannot be
attributed to an unaccounted for broad line component as they are
located well beyond the BLR.

The \oxy kinematic maps (panels h \& i) also show pronounced features
in the off-centre region D.  The observed velocity blob in this region
is offset from the systemic velocity by about 150 km$^{-1}$ and is
located too far from the galaxy centre ($\sim 5 $kpc) to be related to
the kinematics of the nucleus itself.

\subsection{Other Wavelengths}

\name\ has been detected in X-ray and Radio wavelengths at values
typical for a Seyfert galaxy.  Anderson et al. (2003) associate this
system with an X-ray source in the ROSAT All Sky Survey (RASS) of
log$_{10}$($L_X$/erg~s$^{-1})=$~42.86.  From the NVSS 1.4GHz radio
survey we derive a total luminosity of 7.7$\times
10^{22}$W~Hz$^{-1}$. The more accurate position and better resolution
of the FIRST 1.4GHz survey shows that this radio source is centred on
the host galaxy nucleus, coincident with the ENLR.  It appears to be
unresolved in these data implying an upper limit on its projected size
of 7.2 kpc (see Fig.~2 panel~h).

\section{Discussion and Conclusions}\label{sec:discussion}

The high ionization state of the ENLR (with a ratio [OIII]/$H\beta$ in
the range 5-10) is suggestive of gas photo-ionized by the AGN.
Ioniozation from star formation is not so hard, and shocks also tend
to produce lower line ratios (Villar-Mart\'in et al. 1999).  With a
radius along its long axis of $\gtrsim 9$ kpc the ENLR is quite large.
However, larger ENLRs extending up to 20 kpc in Seyfert galaxies have
been observed before (Unger et al. 1987; Fraquelli et al. 2000).  In
order for the AGN ionizing radiation to reach such radii, the AGN
ionization cone must be pointing out of the galaxy disk and excite the
extra-planar gas (Storchi-Bergmann et al. 1992).  The presence of this
gas and its observed high velocity dispersion suggests kinematic
heating by the AGN.

The detailed correspondence between features in radio maps and
emission line images of Sy 2 galaxies suggest strong interactions
between the radio jets and the ENLR (e.g. Falcke et al. 1998;
(Villar-Mart\'in et al. 1999).  Hence, the blob of perturbed
kinematics in \name\ (region D in Fig.~2) at $\sim 5$ kpc from the
nucleus along the ENLR axis is expected to be closely aligned with the
radio structure.  Although we cannot discard alternative scenarios, an
interesting possibility is that the kinematic perturbation has been
triggered by interaction with the radio structures.

In this interpretation the observed kinematics reflect material being
driven out by the jet.  The impact of such an outflow on the
surrounding medium stirs up the gas, presumably by the vortices
trailing the jet shock-front, leading to large (random) gas motions.
A caveat here is that the observed kinematics could also be attributed
to gas falling into the nucleus (possibly as the result of the
interaction with the companion galaxy).  However, in that scenario it
is difficult to explain the high velocity dispersion of the gas.

It is interesting to note that on the opposite side of the nucleus
(region E) neither the H$\alpha$ nor the \oxy kinematic maps show any
remarkable features. The ionization state, however, is similar to
region D.  The H$\alpha$ dispersion map does show a large perturbation
in region F.  Perhaps this is the signature of the counter jet
although it could also be due to interaction with the companion
galaxy.

The total [OIII] luminosity that we derive from the flux distribution
shown in panel (g) is $4.2 \times 10^{40}$ erg~s$^{-1}$ (ignoring the
contribution from region C, which is due to star formation).  This is
equivalent to the H$\alpha$ luminosity of the ENLR, $L_{\rm H\alpha} =
4\times10^{40}$ erg~s$^{-1}$.  Note that this will include the
contribution from the gas disk, which cannot be properly separated
from the highly ionized extra-planar gas.
In comparison, the mechanical power of the jet itself can be estimated
using an empirical conversion from radio luminosity (equation (1) of
Best et al. 2007).  The observed radio flux (table~\ref{t:properties})
implies $L_{\rm mech} = (4 \pm 2)\times10^{42}$ erg~s$^{-1}$, roughly
an order of magnitude more than the energy which is reradiated as
emission lines.
The X-ray luminosity (table~\ref{t:properties}) is larger than both
$L_{\rm mech}$ and the emission line luminosities (H$\alpha$ and
[OIII]) as expected (e.g. Heckman et al. 2004).

If we assume that most of the mechanical energy of the jet is
converted to kinetic energy in the extra-planar gas then we can
compute an upper limit on the mass of ionized hydrogen in the ENLR.
Over the lifetime of the jet, $\lesssim 10^6$ yr (e.g. Sanders 1984),
the upper limit on the available energy in the ENLR is: $\lesssim 4
\times 10^{42}$ erg~s$^{-1} \times \lesssim 10^6$ yr $ \simeq 1.3
\times 10^{56}$ erg.  The jet lifetime is also consistent with its
small size of $\lesssim 5$ kpc, assuming a canonical jet velocity of
$\gtrsim 0.1c$.  With the typical gas velocities observed in our data
of $V^2_{\rm RMS} = V^2_{\rm rot} + \sigma^2 \sim 300$ km~s$^{-1}$,
this kinetic energy would correspond to an upper limit on the mass in
ionized hydrogen of $\sim1.4 \times10^{8} M_\odot$.

In our interpretation, we associate the structure observed in region D
with jet driven mass outflow.  The fraction of the total kinetic
energy needed to power this outflow is simply the fraction of mass in
region D multiplied by the ratio of the bulk velocity ($\sim 150$
km~s$^{-1}$) to $V_{\rm RMS} \sim 300$ km~s$^{-1}$ squared.  Under the
assumption of constant gas density, the fraction of mass can be
estimated as the fraction of [OIII] luminosity in region D, $\sim
0.07$.  Therefore the fraction of ENLR kinetic energy in this bulk
outflow is $\sim 1.3 \times 10^{56} \times 0.07 \times (150/300)^2
\simeq 2.3 \times 10^{54}$ erg.  That is, only about 2 percent of the
mechanical energy is required to power the outflow.

The derived energies are order of magnitude estimates only but are all
internally consistent.  The low mass loading and velocity associated
with the outflow makes it unlikely that this process has a profound
impact on the cold gas content of this galaxy.  However, the implied
mechanical energy of the jet is 50 times greater --- on this basis
only a small fraction of the jet energy is used to power the outflow.
A much larger fraction is available to heat the gas which we observe
as the highly ionized, large ENLR in this galaxy.
It is notable that the jet energy is comparable to the cooling
luminosity of a 1 keV ($\sim 10^{13.5} {\rm M_{\odot}}$) halo. This is
an important point --- in this galaxy feedback from the AGN seems to
have little direct effect on the galaxy: any influence it can have
occurs through the heating of gas in the galaxy's halo.  This scenario
is very much consistent with current galaxy formation models (eg.,
Croton et al.\ 2006, Bower et al.\ 2006, 2008).  Compared with
powerful QSOs (e.g. Nesvadba et al. 2007) and radio galaxies (eg.,
Best et al.\ 2007) the jet energy is small.  Nevertheless, it is the
impact of AGN feedback in $10^{12}$--$10^{13} {\rm M_{\odot}}$ haloes
that is responsible for shaping the galaxy luminosity function.

The jet of this low mass AGN imparts more of its kinetic energy into
the cold gas by means of kinetic heating than by directed outflow.
IFU observations of galaxies hosting radio AGN, such as presented in
this letter, provide key insight into the coupling between the jet and
the gas.

\section*{Acknowledgments}

We thank the referee, Montserrat Villar-Martin, for the constructive
comments and suggestions.  We also like to thank Chris Done, Isabelle
Gavignaud, Martin Krause and Marc Schartmann for helpful discussions.

\label{lastpage}


\begin{thebibliography}{99}

\bibitem[\protect\citeauthoryear{Allen et al.}{2006}]{allen06} 
  Allen, S. W., R. J. H Dunn, Fabian, A. C., Taylor, G. B., \&
  Reynolds, C. S. 2006, MNRAS, 372, 21

\bibitem[\protect\citeauthoryear{Anderson}{2003}]{and03} 
  Anderson, S. F., Voges, W., Morgan, B., Tr\"umper, J., Ag\"ueros, M. A.,
  Boller, T., Collinge, M. J. et al. 2003, AJ, 126, 2209

\bibitem[\protect\citeauthoryear{Baldwin}{1981}]{bal81} 
  Baldwin, J. A., Phillips, M. M., Terlevich, R. 1981, PASP, 93, 5

\bibitem[\protect\citeauthoryear{Bennert}{2006}]{ben06} 
  Bennert, N., Jungwiert, B., Komossa, S., Haas, M., \& Chini, R. 
  2006, A\&A, 456, 953

\bibitem[\protect\citeauthoryear{Best}{2007}]{bes07} 
  Best, P. N., von der Linden, A., Kauffmann, G., Heckman, T. M.,
  \&  Kaiser, C. R. 2007, MNRAS, 379, 894

\bibitem[\protect\citeauthoryear{Bower}{2006}]{bow06} 
  Bower, R. G., Benson, A. J., Malbon, R., Helly, J. C., Frenk, C. S., 
  Baugh, C. M., Cole, S. \& Lacey, C. G. 2006, MNRAS, 370, 645

\bibitem[\protect\citeauthoryear{Bower et al.}{2008}]{bow08} 
   Bower, R. G., McCarthy I. G., Benson A. 2008, MNRAS, submitted

\bibitem[\protect\citeauthoryear{Brinchmann}{2004}]{bri04} 
  Brinchmann, J., Charlot, S., White, S. D. M., Tremonti, C., Kauffmann, G.,
  Heckman, T. \& Brinkmann, J. 2004, MNRAS, 351, 1151

\bibitem[\protect\citeauthoryear{Cappellari}{2004}]{cap04} 
  Cappellari, M. \& Emsellem, E. 2004, PASP 116, 138

\bibitem[\protect\citeauthoryear{Cattaneo}{2005}]{cat05} 
  Cattaneo, A., Combes, F., Colombi, S., Bertin, E. \& Melchior, A.-L.
  2005, MNRAS, 359, 1237

\bibitem[\protect\citeauthoryear{Croton et al.}{2006}]{croton06} 
  Croton, D. J., et al. 2006, MNRAS, 365, 11

\bibitem[\protect\citeauthoryear{Di Matteo}{2005}]{dim05} 
  Di Matteo, T.,  Springel, V. \& Hernquist, L. 2005, Nat, 433, 604

\bibitem[\protect\citeauthoryear{Falcke}{1998}]{fal98} 
  Falcke, H., Wilson, A. S. \& Simpson, C. 1998, ApJ, 502, 199

\bibitem[\protect\citeauthoryear{Fraquelli}{2000}]{fra00} 
  Fraquelli, H. A., Storchi-Bergmann, T. \& Binette, L. 2000, ApJ, 532, 867

\bibitem[\protect\citeauthoryear{Heckman}{2004}]{hec04}
  Heckman, T. M.. Kauffmann, G., Brinchmann, J., Charlot, S., Tremonti, C.,
  \& White, S. D. M. 2004, ApJ, 613, 109

\bibitem[\protect\citeauthoryear{Holt}{2008}]{hol08} 
  Holt, J., Tadhunter, C. N. \& Morganti, R. 2008 arXiv:0802.1444v1

\bibitem[\protect\citeauthoryear{Kuo}{2008}]{kuo08} 
 Kuo, C-Y., Lim, J., Tang, Y-W., \& Ho, P. T. P. 2008 (arXiv:0802.4205v1)

\bibitem[\protect\citeauthoryear{Lawrence}{2007}]{law07} 
  Lawrence, A. et al. 2007, MNRAS, 379, 1599

\bibitem[\protect\citeauthoryear{Letawe}{2008}]{let08} 
  Letawe, Y., Magain, G., Letawe, G., Corbin, F. \& Hutsemekers, D.
  2008 arXiv:0802.1386v1

\bibitem[\protect\citeauthoryear{Nesvadba}{2007}]{nes07} 
  Nesvadba, N. P. H., Lehnert, M. D., De~Breuck, C., Gilbert, A. 
  \& van~Breugel, W. 2007, A\&A, 475, 145

\bibitem[\protect\citeauthoryear{Sanders}{1984}]{san84} 
  Sanders, R. H. 1984, A\&A, 140, 52

\bibitem[\protect\citeauthoryear{Sanchez}{2005}]{san05} 
  S\'anchez, S. F., Becker, T., Garcia-Lorenzo, B., Benn, C. R.,
  Christensen, L., Kelz, A., Jahnke, K. \& Roth, M. M. 2005, A\&A, 429L, 21

\bibitem[\protect\citeauthoryear{Springel}{2005}]{spr05} 
  Springel, V., Di~Matteo, T., \& Hernquist, L. 2005, MNRAS, 361, 776	

\bibitem[\protect\citeauthoryear{Storchi-Bergmann}{1992}]{sto92}
  Storchi-Bergmann, T., Wilson, A. S. \& Baldwin, Jack A. 1992, ApJ, 396, 45

\bibitem[\protect\citeauthoryear{Unger}{1987}]{ung87} 
  Unger, S. W., Pedlar, A., Axon, D. J., Whittle, M., Meurs, E. J. A. 
  \& Ward, M. J. 1987 MNRAS, 228, 671

\bibitem[\protect\citeauthoryear{villar}{1999}]{vil99} 
  Villar-Mart\'in, M., Tadhunter, C., Morganti, R., Axon, D. \& Koekemoer, A.
  1999, MNRAS, 307, 24

\bibitem[\protect\citeauthoryear{York}{2000}]{yor00} 
  York, D. G. et al. 2000, AJ, 120, 1579

\end{thebibliography}
\end{document}